\documentclass{aa}  

\usepackage{graphicx}
\usepackage{txfonts}
\usepackage[justification=centering, small]{caption}
\usepackage[english]{babel}
\usepackage{array}
\usepackage{multirow}
\usepackage{longtable}
\usepackage[usenames]{color}

\newcolumntype{C}{>{\centering\arraybackslash}m{2cm}}
\begin{document}

   \title{The largest metallicity difference in twin systems: High-precision abundance analysis of the benchmark pair Krios \& Kronos}

   \titlerunning{Analysis of Kronos \& Krios using MAROON-X spectra}
   \authorrunning{Miquelarena et al.}

   \author{P. Miquelarena\inst{1,2,6}, C. Saffe\inst{1,2,6}, M. Flores\inst{1,2,6}, R. Petrucci\inst{5,6}, J. Yana Galarza\inst{7}, J. Alacoria\inst{1,6}, M. Jaque Arancibia\inst{3,4}, E. Jofr\'e\inst{5,6}, K. Montenegro Armijo\inst{8} and F. Gunella\inst{1,6}
           }

\institute{Instituto de Ciencias Astron\'omicas, de la Tierra y del Espacio (ICATE-CONICET), C.C 467, 5400, San Juan, Argentina.
         \and Universidad Nacional de San Juan (UNSJ), Facultad de Ciencias Exactas, F\'isicas y Naturales (FCEFN), San Juan, Argentina.
         \and Instituto de Investigación Multidisciplinar en Ciencia y Tecnología, Universidad de La Serena, Raúl Bitrán 1305, La Serena, Chile.       
        \and Departamento de Astronomía, Universidad de La Serena, 1305, La Serena, Chile.
         \and Universidad Nacional de Córdoba, Observatorio Astronómico de Córdoba, Laprida 854, Córdoba X5000BGR, Argentina.
        \and Consejo Nacional de Investigaciones Cient\'ificas y T\'ecnicas (CONICET), Argentina.
        \and The Observatories of the Carnegie Institution for Science, 813 Santa Barbara Street, Pasadena, CA 91101,
USA.
           \and Clínica Universidad de los Andes, Chile, Dirección Comercial.
         }

   \date{Received xx, 2024; accepted xx, 2024}

 
  \abstract
  {}
{We conducted a high-precision differential abundance analysis of the remarkable binary system HD 240429/30 (Krios and Kronos, respectively), whose difference in metallicity is one of the highest detected to date in systems with similar components  ($\sim0.20$ dex). A condensation temperature $T_{C}$ trend study was performed to search for possible chemical signatures of planet formation. In addition, other potential scenarios are proposed to explain this disparity.}
{Fundamental atmospheric parameters ($T_{eff}$, $log\ \textit{g}$, [Fe/H], $v_{turb}$) were calculated using the latest version of the FUNDPAR code in conjunction with ATLAS12 model atmospheres and the MOOG code, considering the Sun and then Kronos as references, employing high-resolution MAROON-X spectra. We applied a full line-by-line differential technique to measure the abundances of 26 elements in both stars with equivalent widths and spectral synthesis taking advantage of the non-solar-scaled opacities to achieve the highest precision.}
{We find a difference in metallicity of $\sim0.230$ dex:  Kronos is more metal rich than Krios. This result denotes a challenge for the chemical tagging method. The analysis encompassed the examination of the diffusion effect and primordial chemical differences, concluding that the observed chemical discrepancies in the binary system cannot be solely attributed to any of these processes. The results also show a noticeable excess of Li of approximately $0.56$ dex in Kronos, and an enhancement of refractories with respect to Krios. A photometric study with TESS data was carried out, without finding any signal of possible transiting planets around the stars. Several potential planet formation scenarios were also explored to account for the observed excess in both metallicity and lithium in Kronos;  none was definitively excluded. While planetary engulfment is a plausible explanation, considering the  ingestion of an exceptionally high mass, approximately $\sim27.8M_{\oplus}$, no scenario is definitively ruled out. We emphasize the need for further investigations and refinements in modelling; indispensable for a comprehensive understanding of the intricate dynamics within the Krios \& Kronos binary system.}
{}


   \maketitle
%

\section{Introduction}

The chemical tagging technique consists in the possibility of identifying co-natal stars that have dispersed into the Galactic disc  based on chemistry alone \citep[e.g.][]{freeman02,casamiquela21}. This idea has been one of the motivations of important surveys such as APOGEE, GALAH, and the \textit{Gaia}-ESO survey
\citep{gilmore12,randich13,desilva15,majewski17}. A fundamental assumption guiding these surveys is that the members of the birth cluster should exhibit a chemically homogeneous composition. This hypothesis was tested using main-sequence and red giant stars in open clusters, reaching an internal coherence in metallicity in the range 0.02$-$0.03 dex \citep[e.g.][]{desilva06,bovy16,liu16,casamiquela20,casamiquela21}. Originally proposed by \citet{andrews18}, wide binaries (100 au $< a <$1 pc) are an ideal sample for studying chemical tagging \citep[e.g.][]{andrews19,kamdar19,hawkins20}. In particular, for the case of binaries with physically similar components, it is possible to reach the highest possible precision through a line-by-line differential analysis \citep[e.g.][]{schuler11,saffe15,saffe17,teske16,liu18,tuccimaia19,jofre21,flores23}, which helps to minimize a number of model-induced and other systematic errors \citep[see][]{nissen18}.

Recently, the internal coherence of the chemical tagging was strongly challenged by the discovery of the  exceptional comoving pair HD 240429/30 \citep[hereafter Krios \& Kronos;][]{oh18}, composed of two G-type stars sharing nearly identical \textit{Gaia} TGAS\footnote{\tiny The Tycho-\textit{Gaia} Astrometric Solution catalogue (TGAS) is a component of \textit{Gaia} DR1 \citep{michalik15}.} proper motions and parallaxes. \citet{oh18} suggest that the two stars are co-natal, based on their proximity in phase-space, with very similar radial velocites
and isochrone ages, and also with very low probabilities of stellar capture and exchange scattering. The authors used the stellar parameters and abundances from the survey of \citet{brewer16}, who studied 1617 FGK stars that belong to the California Planet Survey (CPS) using an automated spectral synthesis procedure. In this way, \citet{oh18} estimated for the pair a mutual difference in iron content of $\sim$0.20 dex, and a similar value for other metals such as Ca and Ni. To our knowledge, this is the largest difference found to  date between stars with twin components and a supposed common origin, highlighting the pair Kronos \& Krios as a benchmark multiple system.

\citet{hawkins20} studied 25 binary systems and found that 80\% are homogeneous at the 0.02 dex level, while six pairs show differences greater than 0.05 dex. Then, if confirmed, the metallicity difference between Kronos \& Krios would be ten times higher, in logaritmic scale, than the typical internal
coherence of stars born in the same cluster. The greatest difference found between Kronos \& Krios would imply that their co-natal nature could not be recovered  by any previous chemical tagging work \citep[e.g.][]{desilva06,bovy16,liu16,casamiquela20,casamiquela21}.  The difference between Kronos \& Krios ($\sim$0.2 dex) is similar to those found between random pairs \citep[scatter of 0.23 dex, ][]{nelson21}, defying the main assumption of the chemical tagging, in which stars formed together display the same abundances along their main sequence lifetimes. Recently, \citet{saffe24} analysed, for the first time, a giant-giant binary system bringing new insights, with significant differences in metallicity potentially attributed to primordial inhomogeneities. The significance of these   findings underscores the importance of our binary system and deserves particular attention.

In addition, it is equally important to explain the origin of the significant metallicity difference between Kronos \& Krios. This requires   studying the relative volatile-to-refractory content between the stars and the condensation temperature (T$_{c}$) trends. For instance, \citet{melendez09} found that the Sun is deficient in refractory elements (T$_{c}$ $>$ 900 K) relative to volatile (T$_{c}$ $<$ 900 K) when compared to 11 solar twins, and that the abundance differences correlate with T$_{c}$. They suggested that this trend is a signature of planet formation, assuming that refractory elements were locked up in rocky planets during the Solar System formation. However, different explanations for the T$_{c}$ trends could also be possible. \citet{booth-owen20} suggest that if a giant planet forms early enough ($\la$1 Myr) at large separations, it could trap $\ga$100 M$_{\oplus}$ of dust exterior to its orbit. Then, the star would accrete more gas than dust from the protoplanetary disc, which could result in a lack of refractories in the stellar atmosphere. A larger amount of refractories in a stellar atmosphere could also be the result of accretion of rocky material \citep[e.g.][]{gonzalez97,melendez17,saffe17,oh18}.  \citet{oh18} suggested that Kronos accreted $\sim$15 M$_{\oplus}$ of rocky material in order to explain the mutual T$_{c}$ trend. To date, this is the highest amount of material estimated to be accreted in binary systems with twin components: it is equivalent to approximately seven times the four inner planets
of the Solar System together, which is also remarkable. Other authors consider alternative scenarios trying to explain the T$_{c}$ trends, such as Galactic chemical evolution (GCE) or dust-cleansing effects \citep[e.g.][]{onehag11,adibekyan14,nissen15}.

Recently, \citet{spina21}  studied a sample of 107 binary systems and showed that accretion events occur in $\sim$20-35\% of solar-type stars. In contrast, \citet{behmard23} found a much lower engulfment rate of $\sim$2.9\%, claiming that accretion events are rarely detected. The last authors propose that primordial inhomogeneities rather than engulfment events could explain the differences observed in binary systems. According to their criteria (see Sect. 6), Kronos \& Krios would be the only pair showing a true engulfment detection, ruling out most previous claims of engulfment events. This highlights again the relevance of the notable pair Kronos \& Krios between other binary systems. Interestingly, \citet{kunimoto18} consider  an engulfment event unlikely in this binary system, owing to the rapid mixing expected from fingering convection \citep[10-100 Myr, ][]{tv12}. Thus, the origin of the extreme metallicity difference in this benchmark pair remains unknown.

A number of recent works studied atomic diffusion effects on main-sequence stars, using stellar evolution models \citep[e.g.][]{dotter17},
observing the stars of the M67 open cluster \citep[e.g.][]{souto18,souto19} and also using binary stars \citep[e.g.][]{ramirez19,liu21}.
Diffusion models show a strong dependence on log g, with the largest effects occurring near log g$\ \sim$ 4.2 dex \citep[see e.g. Fig. 5 in][]{souto19}. The same plot predicts that a difference of $\sim$ 0.20 in log\ \textit{g} could translate into a difference of $\sim$ 0.075 in [Fe/H]; other differences are predicted for different chemical elements. \citet{liu21} found that the overall abundance offsets in four of seven binary systems could be due to atomic diffusion effects,
complicating the chemical tagging. The difference in log g estimated for the pair Kronos--Krios is 0.10 dex \citep{brewer16}, the largest difference found in the sample
of twin-star binary systems of \citet{ramirez19}. Then, we wondered if atomic diffusion effects not previously studied in this benchmark pair could explain, at least in part, the extreme difference in metallicity found.

The detection of a possible T$_{c}$ trend in a binary or multiple system is a challenge, requiring the highest possible precision in the derivation of stellar parameters and abundances. This demands  high-quality spectra with very high S/N, reaching typically $\sim$400 or even more \citep[e.g.][]{teske16,liu18,schuler11,tuccimaia19}, compared to S/N$\sim$200 for the case of Kronos \& Krios \citep{oh18}. For stars with low rotational velocities, it is usual to use equivalent widths rather than spectral synthesis in the derivation of stellar parameters, given that spectral synthesis depends on additional factors (such as v $\sin$ i, the resolving power R of the instrument, and the correct fitting of line profiles). The stars Kronos \& Krios present projected rotational velocities of 1.1 km/s and 2.5 km/s \citep{brewer16}, allowing  a clean measurement of equivalent widths.
Moreover, for the case of multiple systems with physically similar components, the use of a line-by-line differential technique allows the minimization of systematic errors \citep[e.g.][]{schuler11,bedell14,saffe15,teske16,liu18,tuccimaia19}. In this way, the physical similarity between Kronos \& Krios (G0V+G2V) is an advantage to be exploited
with a differential analysis, a technique not applied by previous works for this pair.

Then we studied the benchmark pair Kronos \& Krios by using a high-quality MAROON-X spectra with higher S/N ($\sim$400), higher resolving power (R $\sim$ 85000), broader spectral coverage (from ${\sim}4900$ to $9200{\AA}$), and using a more refined analysis technique than previous works (fully differential together with equivalent widths). In addition, we took advantage of using non-solar-scaled opacities in the derivation of model atmospheres, which could result in small abundance differences when compared to the classical solar-scaled methods \citep{saffe18,saffe19,flores23}. This  allowed us to determine a metallicity difference between the two stars with the highest possible precision, and to perform a T$_{c}$ trend analysis to study the possible origin of the differences in this benchmark pair, which could be attributed to a planet engulfment event \citep{oh18,behmard23}. Moreover, we explored alternative scenarios that could lead to this result, such as atomic diffusion \citep{liu21} and the potential primordial origin of the chemical difference \citep{ramirez19,nelson21,saffe24}.

This work is organized as follows. In Sect. 2 we describe the observations and data reduction. In Sect. 3 we present the stellar parameters and chemical abundance \mbox{analysis}. In Sect. 4 we show the results and discussion. Finally, in Sect. 5 we highlight our main conclusions.


\section{Observations and data reduction}

The spectra of Kronos \& Krios were acquired through the M-dwarf Advanced Radial velocity Observer Of Neighboring eXoplanets (MAROON-X) spectrograph.\footnote{\tiny \url{https://www.gemini.edu/instrumentation/maroon-x}} This high-precision bench-mounted echelle spectrograph provides  high-resolution (R $\sim$ 85000) spectra when illuminated via
two 100 $\mu$m (0."77 on sky) octogonal fibres. MAROON-X is connected to the 8.1 m Gemini North telescope at Maunakea, Hawaii. Currently, the spectrograph has no movable parts and is operated in one read-out mode (100 KHz, 1x1 binning). MAROON-X is equiped with two STA4850 (4080 x 4080) CCD detectors with a pixel size of 15 $\mu$m, including a coating optimized for their respective wavelength coverage. The instrument includes its own tungsten-halogen lamp for flat-fielding and a ThAr arc lamp for wavelength calibration.

The observations were taken on August 15, 2022 (Programme ID: GN-2022B-Q-203, PI: Paula Miquelarena); the star Kronos was observed immediately after the star Krios, using the same spectrograph configuration. The exposure times for Krios and Kronos were 3 x 20 minutes and 3 x 16.67 minutes, respectively. This resulted in a final signal-to-noise ratio (S/N) per pixel of $\sim$420 for both stars, measured near $\sim$6000 {\AA} in the combined spectra. The final spectral coverage was ${\sim}4900-9200{\AA}$. The solar spectrum was obtained by observing the asteroid Vesta (Programme ID: GN-2022A-Q-22, PI: Yuri Netto), yielding a S/N similar to that achieved in the combined spectra of Kronos and Krios. However, it is worth mentioning that the most accurate differential study, in terms of abundance precision, is conducted between the components of the binary system due to their similarity.

MAROON-X spectra were reduced using MAROONXDR,\footnote{\tiny \url{https://github.com/GeminiDRSoftware/MAROONXDR}} a publicly available Data Reduction for Astronomy from Gemini Observatory North and South \citep[DRAGONS, ][]{labrie19} implementation of the data reduction pipeline, following the standard recipe for echelle spectra (e.g. bias and flat corrections, scattered light correction). The continuum normalization and other operations (such as Doppler correction and spectra combination) were carried out using the Image Reduction and Analysis Facility (IRAF).\footnote{ \tiny IRAF is distributed by the National Optical Astronomical Observatories, which is operated by the Association of Universities for Research in Astronomy, Inc. (AURA), under a cooperative agreement with the National Science Foundation.}

\begin{table*}
    \centering
    \renewcommand{\thefootnote}{\thempfootnote}
    \captionsetup{justification=justified,singlelinecheck=false} 
     \caption{Fundamental parameters obtained for Kronos and Krios.}
    \begin{tabular}{cccccccc}
    \hline
    \noalign{\smallskip}
    \multicolumn{1}{c}{\multirow{2}{*}{Star}} & T$_{\rm eff}$ & $\log g$ & [Fe/H] & v$_\mathrm{micro}$ & T$_{\rm eff}^\dagger$ & $\log g^\star$\\
        &  \small [K]          &  \small [dex]   & \small [dex]      & \small [km s$^{-1}$] &  \small [K]          &  \small [dex]    \\
     \noalign{\smallskip} 
    \hline
    \noalign{\smallskip}
    \multicolumn{7}{c}{\textbf{Our work}} \\
    \hline
    \noalign{\smallskip}
    Kronos - Sun  &   5895 $\pm$ 66   &   4.44 $\pm$ 0.06  &   0.220 $\pm$ 0.007  &   1.18 $\pm$ 0.04  &  5903 $\pm$ 51  &  4.40 $\pm$ 0.04      \\
    Krios - Sun   &   5892 $\pm$ 52    &   4.49 $\pm$ 0.08  &-0.010 $\pm$ 0.010   &   1.18 $\pm$ 0.06  &  5938 $\pm$ 40 &  4.45 $\pm$ 0.04   \\
    Krios - Kronos   &   5895 $\pm$ 38    &   4.49 $\pm$ 0.05  &-0.230 $\pm$ 0.005   &   1.19 $\pm$ 0.04 & -  & -  \\
    \noalign{\smallskip}
    \hline
    \noalign{\smallskip}
     \multicolumn{7}{c}{\textbf{\citet{brewer16}}} \\
    \hline
    \noalign{\smallskip}
    Kronos - Sun  &   5803 $\pm$ 25        &   4.33 $\pm$ 0.03  &   0.20 $\pm$ 0.010  &   0.85 & -  &  -    \\
    Krios - Sun   &   5878 $\pm$ 25     &   4.43 $\pm$ 0.03  & 0.01 $\pm$ 0.010   &   0.85  & -  & - \\
    \noalign{\smallskip}
    \hline
\end{tabular}
\tablefoot{
\tablefoottext{$^\dagger$}{Photometric effective temperature.}
\tablefoottext{$^\star$ }{Trigonometric $\log g$ obtained from PARAM 1.5.}

}
\label{table.params}
\end{table*}

\section{Stellar parameters and abundance analysis}

We determined fundamental stellar parameters, such as effective temperature ($T_{eff}$), surface gravity (log $g$), metallicity ($[Fe/H]$), and microturbulence velocity ($v_{turb}$), as well as chemical abundances for Kronos and Krios by first measuring the equivalent widths (EWs) of 26 elements, including \ion{Fe}{I} and \ion{Fe}{II}, using the \textit{splot} task in IRAF. The list of spectral lines, along with significant laboratory data, such as excitation potential, oscillator strengths, and log \textit{gf} values, were sourced from \citet{liu14}, \citet{melendez14}, and were supplemented with data from \citet{bedell14}, who carefully selected lines for precise abundance determinations.

\begin{figure}
\centering
\includegraphics[width=9cm]{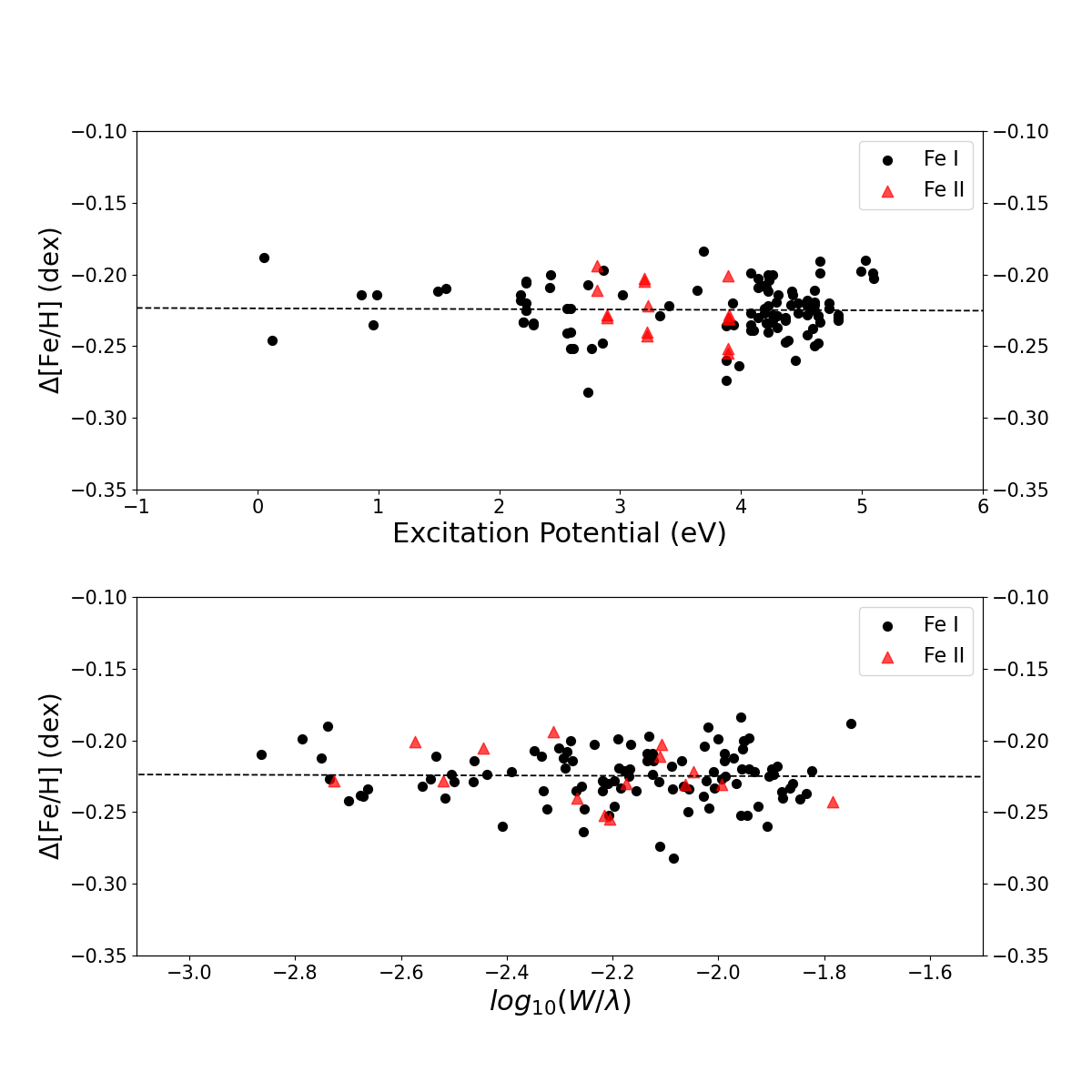}
\captionsetup{justification=justified,singlelinecheck=false} 
\caption{Differential abundance vs excitation potential (upper panel) and differential abundance vs reduced EW (lower panel) of Krios relative to Kronos. The black dots correspond to \ion{Fe}{I} and the red triangles correspond to \ion{Fe}{II}.}
\label{eq.krios.kronos}%
\end{figure}

Stellar atmospheric parameters were obtained by imposing ionization and excitation balance of the  \ion{Fe}{I} and \ion{Fe}{II} lines. In this method we search for a zero slope when comparing \ion{Fe}{I} and \ion{Fe}{II} abundances with reduced equivalent width ($EW_{r}=EW/\lambda$) and excitation potential, respectively. For this purpose, we employed the FUNdamental PARameters programme \citep[FUNDPAR,][]{saffe15,saffe18} in its latest version. It uses the  MOOG\footnote{\tiny \url{https://www.as.utexas.edu/ \sim chris/moog.html}} code \citep{sneden73} together with ATLAS12 model atmospheres \citep{kurucz93} to search for the best solution \citep[for more details, see][]{saffe18}. In Figure \ref{eq.krios.kronos} we present the differential abundances of \ion{Fe}{I} (black) and \ion{Fe}{II} (red) versus excitation potential (upper panel) and reduced EWs (lower panel) for Krios compared to Kronos.

We employed a full\footnote{\tiny We take into account line-by-line level variations, not only for determining abundances, but also in the calculation of stellar parameters.} line by line differential technique using the Sun as reference in the first step. In this context, the adopted solar parameters were $T_{eff}=5777$ K, $log\ \textit{g}=4.44$ dex, $[Fe/H]=0.00$ dex and $v_{turb}=1.00$ kms$^{-1}$. Subsequently, we recalculated $v_{turb}$ by ensuring a zero slope between absolute abundances of \ion{Fe}{I} and $EW_{r}$, and the value obtained was $1.13$ km s$^{-1}$. The final parameters for Kronos and Krios relative to the Sun are presented in Table \ref{table.params}. The corresponding uncertainities were estimated using the method described in \citet{saffe15}, which accounts for the individual and mutual co-variances for the error propagation. 
We applied the same methodology to determine the differential stellar parameters and abundances of Krios, using Kronos as the reference star. The resulting parameters for Krios relative to Kronos are also provided in Table \ref{table.params}. 

We also derived chemical abundances for 26 elements, other than Fe: \ion{Li}{I}, \ion{C}{I}, \ion{O}{I}, \ion{Na}{I}, \ion{Mg}{I}, \ion{Al}{I}, \ion{Si}{I}, \ion{S}{I}, \ion{Ca}{I}, \ion{Sc}{I}, \ion{Sc}{II}, \ion{Ti}{I}, \ion{Ti}{II}, \ion{V}{I}, \ion{Cr}{I}, \ion{Cr}{II}, \ion{Mn}{I},  \ion{Co}{I}, \ion{Ni}{I}, \ion{Cu}{I}, \ion{Zn}{I}, \ion{Y}{II}, \ion{Zr}{II}, \ion{Ba}{II}, \ion{La}{II}, \ion{Ce}{II}, \ion{Pr}{II}, \ion{Nd}{II}, and \ion{Eu}{II}. For this purpose we implemented a curve of growth analysis by using the latest version of MOOG \citep{sneden73}. In order to account for hyperfine structure (HFS) effects, we employed spectral synthesis for \ion{V}{I}, \ion{Mn}{I},  \ion{Co}{I}, \ion{Cu}{I}, \ion{Li}{I}, \ion{Y}{II}, \ion{Sc}{II}, and \ion{Eu}{II}, incorporating HFS constants from \citet{kurucz-bell95}. We also applied abundance corrections for galactic chemical evolution (GCE) based on the [X/Fe]--age correlation from \citet{bedell18} for (Krios-Sun) and (Kronos-Sun), following the methodology detailed by \citet{spina16} and \citet{yana16-2}. No GCE correction was made for Krios--Kronos, as it is assumed that they were born from the same molecular cloud. Specifically, we considered non-local thermodynamic equilibrium (NLTE) corrections for \ion{Ba}{II} \citep{korotin11},  \ion{Na}{I} \citep{shi04}, and \ion{O}{I} \citep{ramirez07}. The NLTE correction for \ion{Ba}{II} is +0.015 dex for Kronos and 0.00 dex for Krios. For \ion{Na}{I} we adopted -0.08 dex for both stars, and for \ion{O}{I} we adopted +0.11 dex for Kronos and +0.18 dex for Krios. The differential abundances of all elements, along with their corresponding errors, are detailed in Table \ref{table.abund}. It is worth  mentioning that extensive NLTE corrections are avilable using an interpolation tool at the MPIA website.\footnote{\url{https://nlte.mpia.de}} This service includes several elements (Mg, Si, and Ca, among others) and also \ion{Fe}{I} and \ion{Fe}{II} corrections. For example, the \ion{O}{I} triplet include hydrogen collisions with cross-sections based on quantum-mechanical calculations \citep{bergemann21}. The interpolation tool made use of MAFAGS or MARCS model atmospheres. Considering that our calculation used the  ATLAS12 model, a future implementation of FUNDPAR using the MARCS models could take advantage of the mentioned NLTE corrections. The total abundance errors ($\sigma_{TOT}$) were obtained by quadratically adding the observational errors (derived as $\sigma / \sqrt{(n-1)}$) and errors due to uncertainties in fundamental parameters. For those elements with only one line, we adopted for $\sigma$ the average standard deviation of the other elements.

Using the spectroscopic stellar parameters obtained for both components, we derived new values for stellar masses $M_{\star}$, radius $R_{\star}$, and ages $\tau_{\star}$. To accomplish this, we employed PARAM 1.5\footnote{\tiny \url{http://stev.oapd.inaf.it/cgi-bin/param/}}  from the PAdova and tRieste Stellar Evolution Code (PARSEC) \citep{desilva06, rodrigues14, rodrigues17}.
 We specifically utilized the evolutionary tracks from Modules for Experiments in Stellar Astrophysics \citep[MESA;][]{paxton11, paxton13, paxton15, paxton18}; the initial data required for the analysis included $T_{eff}$, $log g$, and [Fe/H], along with the respective $1\sigma$ error in all cases. We also included parallaxes from \textit{Gaia} EDR3 \citep{gaia21} and photometry from Tycho-2 catalogue in V and B bands  \citep{hog00}. The derived values are $M_{\star}= 1.14 \ ^{+ 0.02}_{-0.04}\ M_{\odot} $, $R_{\star}= 1.11 \ ^{+ 0.06}_{-0.04}\ R_{\odot}$, $\tau_{\star}=1.95 \ ^{+ 1.68}_{-1.34}\ Gyr$  for Kronos, and $M_{\star}=1.04 \ ^{+ 0.02}_{-0.02}\ M_{\odot}$, $R_{\star}=1.01 \ \pm 0.04\ R_{\odot}$, $\tau_{\star}=1.57 \ ^{+ 1.67}_{-1.10}\ Gyr$  for Krios. In addition, we estimated the ages of the components using trigonometric log $g$, obtaining as a result $\tau_{\star}=2.18 \ \pm 1.37\ Gyr$ for Kronos and $\tau_{\star}=2.09 \ \pm 1.50\ Gyr$ for Krios, which are similar to the previous values within the errors, providing evidence of the true coevality of the system.

\section{Results and discussion}

The stellar parameters and chemical abundances derived from this work were obtained through the opacity sampling method, incorporating non-solar-scaled opacities \citep{saffe18}. When comparing the fundamental atmospheric parameters listed in Table \ref{table.params} with those obtained from \citet{brewer16}, we find a good agreement within the errors. However, a notable discrepancy arises when comparing $T_{eff}$ differences between the two components. In our investigation these temperatures exhibit notable similarity, yielding identical temperatures when using Kronos as the reference star. In contrast, \citet{brewer16} reports a significant temperature difference between the components. We attribute this discrepancy to the use of higher S/N spectra, the use of different line lists and atmospheric models, and the full line-by-line differential technique employeed in our study. 

Moreover, it is noteworthy that the atmospheric model utilized in the prior chemical analysis, as indicated by \citet{brewer16}, employed a fixed microturbulence parameter set at $0.85\ kms^{-1}$. \citet{nissen18} have cautioned against the potential inaccuracies associated with using a constant value for $v_{turb}$. This caution gains particular significance considering an observed variation of approximately $1.2\ kms^{-1}$ when analysing stars with effective temperatures ranging between 5000 K and 6500 K \citep{edvardsson93, ramirez13}. In our study, we opted not to fix $v_{turb}$; instead, we estimated the value that best fits with the atmosphere model of the components, achieving an optimal agreement between abundances and line intensity. 

Additionally, we calculated the photometric temperatures of the two stars using the  \textsc{colte} code,\footnote{\tiny \url{https://github.com/casaluca/colte}} which derives colour-effective temperature relations employing \textit{Gaia} DR3 and 2MASS photometry in the InfraRed Flux Method, and estimating errors from Monte Carlo simulations of each index \citep{casagrande21}. The weighted average results can be observed in Table \ref{table.params}. For Kronos there is excellent concordance between spectroscopic and photometric $T_{eff}$, and for Krios the photometric temperature appears marginally higher than the spectroscopic value, although still statistically indistinguishable within the errors. Nevertheless, the spectroscopic estimate exhibits a slightly closer agreement with the photometric value compared to those derived by \citet{brewer16}.

\begin{figure}[h!]
        \centering
        \includegraphics[width=9cm]{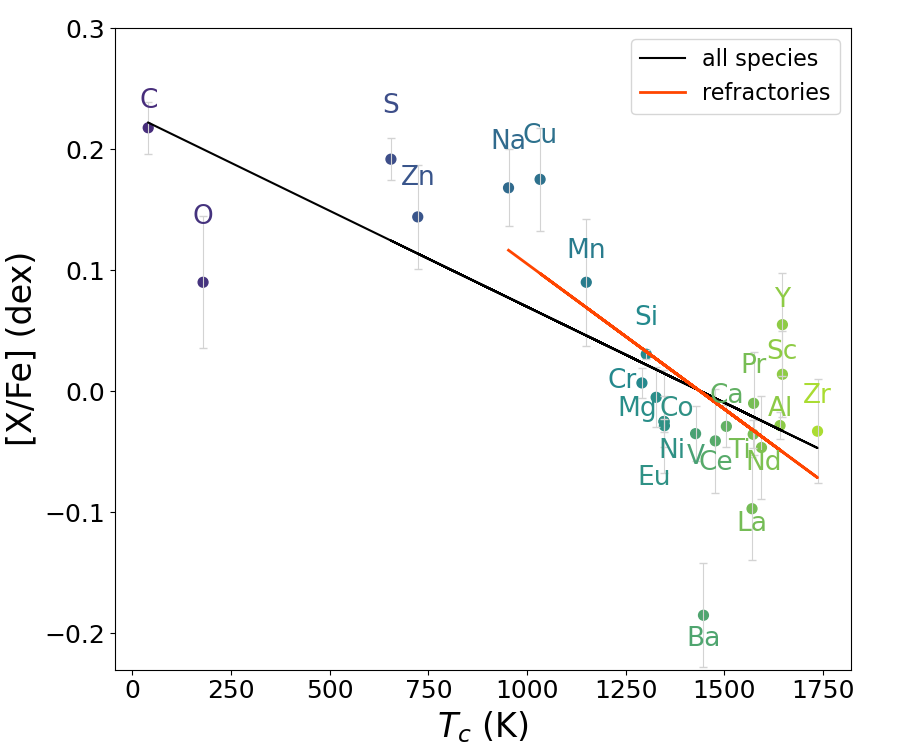}
        \captionsetup{justification=justified,singlelinecheck=false} 
        \caption{Differential abundances from Krios-Kronos vs $T_{C}$. The weighted linear fits to all elements and to refractories are represented as a black and a red line, respectively.}
        \label{tcond}%
\end{figure}

The significant difference in metallicity found in \citet{oh18} of $\sim\ 0.20$ dex is also reflected in this study, with a difference of 0.230 dex, indicating that Kronos is more metal-rich than Krios.
Figure \ref{tcond} shows the abundance of chemical elements in Krios versus condensation temperature $T_{c}$, considering Kronos as reference. The 50\% $T_{C}$ values were taken from \citet{lodders03}, for a solar composition gas. We calculated the slope considering all elements and considering only the refractories. The weighted results were -17.43 $\pm$ 2.25 $\times 10^{-5}$ dex $K^{-1}$ for all elements and -23.98 $\pm$ 5.16 $\times 10^{-5}$ dex $K^{-1}$ for refractories. Based on these findings, a pronounced lack of refractories relative to volatiles in Krios compared to Kronos is evident, with a significance at a $9\sigma$ level. Regarding the refractory elements, we can observe that this slope is also significant at a $6\sigma$ level.

\begin{table}[h!]
\begin{minipage}{\columnwidth}
\renewcommand{\thefootnote}{\thempfootnote}
\captionsetup{justification=justified,singlelinecheck=false} 
\caption{Differential abundances obtained for Kronos and Krios relative to the Sun and for Krios relative to Kronos.}

\scalebox{0.87}{\begin{tabular}{lrrrrrrrrr} 
\cline{1-9}
\noalign{\smallskip}
\multicolumn{1}{c}{\multirow{2}{*}{Species}}   & \multicolumn{2}{c}{Kronos-Sun}               &                      & \multicolumn{2}{c}{Krios-Sun}              &                      & \multicolumn{2}{c}{Krios-Kronos}            &                      \\ 
\noalign{\smallskip} 
\cline{2-3} \cline{5-6} \cline{8-9}
 \noalign{\smallskip}
          & {[}X/Fe{]}           & $\sigma_{TOT}$                &                      & {[}X/Fe{]}           & $\sigma_{TOT}$                  &                      & {[}X/Fe{]}           & $\sigma_{TOT}$                  &                      \\ 
\noalign{\smallskip}
 \cline{1-9}
 \noalign{\smallskip}

 C  I & -0.295 & 0.034 & & -0.070 & 0.040 &  & 0.218 & 0.029\\
  O  I & -0.259 & 0.079 & & -0.163 & 0.062 & & 0.090 & 0.055\\
  Na I & -0.271 & 0.046 & & -0.101 & 0.031 & & 0.168 & 0.030\\
  Mg I & 0.016 & 0.067 & & 0.014 & 0.089 & & -0.005 & 0.044\\
  Al I & 0.075 & 0.018 & & 0.051 & 0.029 & & -0.028 & 0.013\\
  Si I & -0.017 & 0.006 & & 0.016 & 0.023 & & 0.031 & 0.016\\
  S  I & -0.235 & 0.032 & & -0.037 & 0.026 & & 0.192 & 0.024\\
  Ca I & 0.050 & 0.026 & & 0.020 & 0.039 & & -0.029 & 0.019\\
  Sc I & 0.013 & 0.054 & & 0.027 & 0.065 & & 0.014 & 0.053\\
  Sc II & 0.060 & 0.039 & & 0.011 & 0.022 & & -0.059 & 0.02\\
  Ti I & 0.054 & 0.017 & & 0.017 & 0.018 & & -0.036 & 0.013\\
  Ti II & 0.047 & 0.034 & & 0.037 & 0.032 & & -0.014 & 0.027\\
  V  I & 0.023 & 0.032 & & -0.008 & 0.035 & & -0.036 & 0.023\\
  Cr I & -0.019 & 0.019 & & -0.014 & 0.020 & & 0.007 & 0.014\\
  Cr II & -0.021 & 0.028 & & 0.002 & 0.032 & & 0.021 & 0.027\\
  Mn I & -0.169 & 0.081 & & -0.078 & 0.073 & & 0.090 & 0.065\\
  Co I & 0.029 & 0.029 & & 0.008 & 0.031 & & -0.025 & 0.021\\
  Ni I & 0.023 & 0.008 & & -0.004 & 0.011 & & -0.028 & 0.006\\
  Cu I & -0.251 & 0.064 & & -0.070 & 0.050 & & 0.175 & 0.045\\
  Zn I & -0.108 & 0.065 & & 0.041 & 0.062 & & 0.144 & 0.052\\
 Y II & -0.083 & 0.062 & & -0.037 & 0.065 & & 0.055 & 0.052\\
  Zr II & 0.051 & 0.065 & & 0.010 & 0.062 & & -0.033 & 0.052\\
  Ba II & 0.256 & 0.065 & & 0.059 & 0.062 & & -0.185 & 0.052\\
  La II & 0.032 & 0.065 & & -0.075 & 0.062 & & -0.097 & 0.052\\
  Ce II & -0.067 & 0.065 & & -0.118 & 0.062 & & -0.041 & 0.052\\
  Pr II & -0.013 & 0.065 & &-0.029 & 0.062 &  &-0.010 & 0.052\\
  Nd II & 0.196 & 0.051 & & 0.142 & 0.067 & & -0.047 & 0.052\\
  Eu II & 0.144 & 0.065 & & 0.118 & 0.062 & & -0.025 & 0.052\\ \cline{1-9} \noalign{\smallskip}\noalign{\smallskip}
  A(Li)\tablefootmark{a} &2.84  & 0.070 & & 2.28 & 0.070 & & -0.56 &0.069& \\

\end{tabular}}
\tablefoot{
The total error $\sigma_{TOT}$ includes errors due to parameters and observational errors.\\
\tablefoottext{a}{Absolute abundance of Li.}
}
\normalsize
\label{table.abund}
        \end{minipage}
\end{table}

In view of their results, \citet{oh18} explored the possibility that this system formed through binary-single scattering events, where initially unrelated stars undergo an exchange of binary members. The study delves into the rate of exchange scattering, considering factors such as the cross-section and velocity parameters. However, the analysis revealed that this mechanism is unlikely to explain the distinctive abundance patterns observed in such stars. A statistical examination, employing randomly drawn star pairs with similar metallicity characteristics, reinforces this conclusion, highlighting the improbable nature of exchange scattering in accounting for the observed chemical differences within the binary system. 

\subsection{Li content in Kronos \& Krios}
\label{sec:Li}

The Li abundance was initially calculated for Krios and Kronos using spectral synthesis of the $6707.8\ \AA$ line and corrected for NLTE effects using the INSPECT tool \citep{lind12}, obtaining A(Li)=$2.78 \pm 0.07$ dex for Kronos and A(Li)=$2.26 \pm 0.07$ dex for Krios. However, due to an artefact observed around the lithium line, particularly on its left wing, we opted to use spectra from the HIRES database (Programme ID: Y219, PI: Brewer) to redetermine its abundance. After correcting for NLTE effects using the INSPECT tool, we obtained A(Li) values of $2.84 \pm 0.07$ dex for Kronos and $2.28 \pm 0.07$ dex for Krios, in good agreement with the values obtained with MAROON-X spectra, within the errors. Consequently, the lithium difference between components is $\Delta (Li) = 0.56$ dex, slightly greater than the $\Delta (Li) = 0.50$ dex reported by \citet{oh18}.

Prior studies of FGK dwarf and subgiant stars revealed a subtle trend between lithium abundance and $T_{eff}$, with A(Li) being higher for hotter stars \citep{ramirez12, bensby18}. Furthermore, \citet{carlos19} found a strong correlation between Li depletion and age for a sample of 77 solar-type stars, and a weaker correlation with metallicity and mass, with higher Li depletion for older, more metallic, and less massive stars, in line with previous studies \citep[e.g.][]{castro09, carlos16}.

\begin{figure}[h!]
        \includegraphics[width=9.5cm]{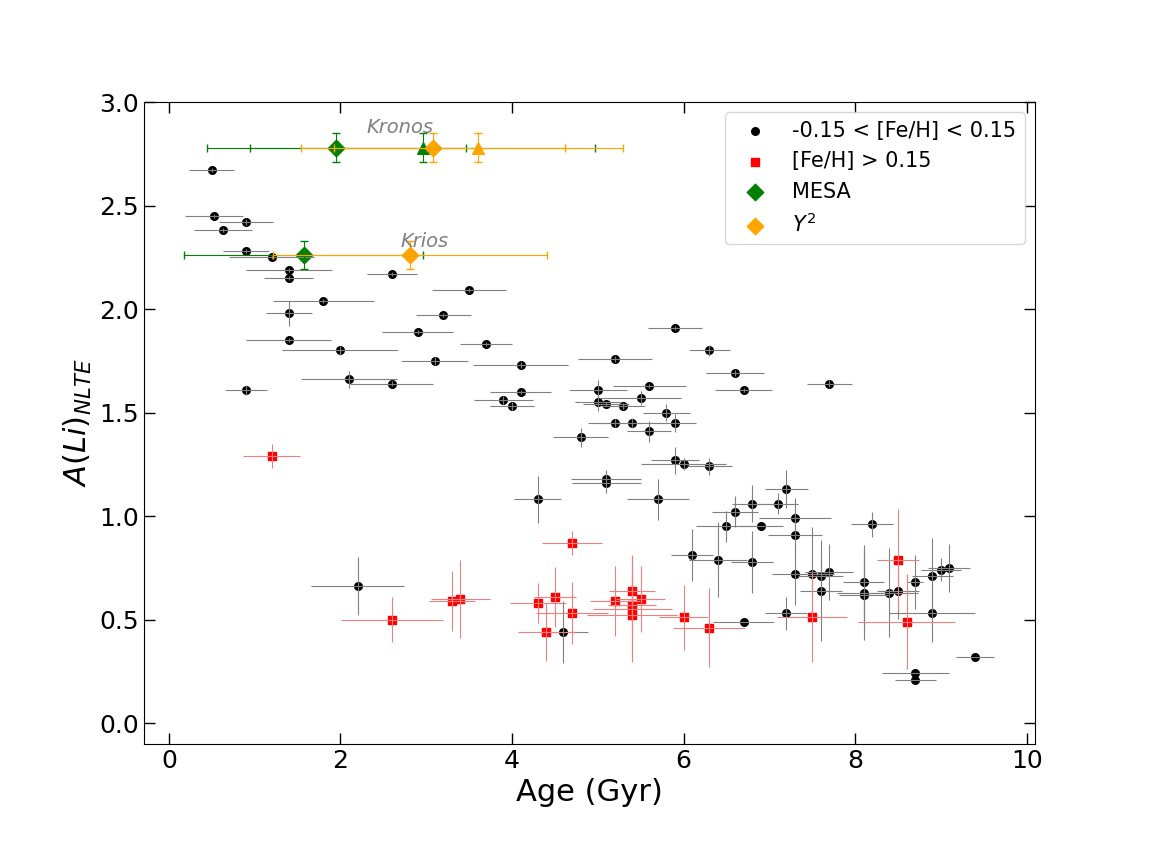}
        \captionsetup{justification=justified,singlelinecheck=false} 
        \caption{Lithium abundance vs age for a sample of solar analogues extracted from \citet{martos23}. The orange diamonds represent Kronos and Krios with ages calculated using $Y^{2}$ isochrones, along with their respective metallicities. The orange triangle represents Kronos with a bulk metallicity composition of [Fe/H]=$-0.01$ dex. Similarly, the green diamonds and triangle represent Kronos and Krios, considering ages calculated with MESA isochrones.}
        \label{martos}%
\end{figure}

 Recently, \citet{martos23} estimated a correlation between Li abundance and both age and [Fe/H] in a sample of 118 solar analogues, using a least-squares method, and found a robust anticorrelation with these parameters. In Figure 3 of their work, they showed the behaviour of A(Li) with respect to Age and [Fe/H]. In Figure \ref{martos}, we replicated this distribution by plotting A(Li)$_{NLTE}$ versus age, including those objects with $-0.15<[Fe/H]<0.15$ (black points) and $[Fe/H]>0.15$ (red squares). We   included Kronos and Krios, shown in the figure with diamonds. Given the significant difference in metallicity between both stars, we also contemplated the hypothesis that the bulk composition of Kronos closely resembled that of Krios, indicated with triangles in the figure. We considered ages computed using MESA isochrones, indicated in green in Figure \ref{martos}. Additionally, we incorporated ages calculated through the Yonsei-Yale ($Y^{2}$) set of isochrones \citep{yi01,demarque04} and taking into account the influence of alpha enhacenment, to maintain consistency with the sample analysed by \citet{martos23}, resulting in  $\tau_{\star}=3.08 \pm 1.54\ Gyr$ and $\tau_{\star}=2.81 \pm 1.60\ Gyr$ for Kronos and Krios, and $\tau_{\star}=3.61\pm 1.69\ Gyr$ for Kronos considering $[Fe/H]=-0.01\ dex$, represented in the figure in orange. First, we focussed on the MESA set of parameters; it is apparent that Krios has a similar A(Li) to the other stars in the same age group. However, the behaviour of Kronos is quite different from the stars of the same age and metallicity in the sample. It can be observed that, regardless of the primordial metallicity that  Kronos may have had, it has more lithium than the rest of the stars in the sample. This phenomenon remains prominent in both cases, whether its primordial metallicity was [Fe/H] = -0.01 dex initially, or considering a bulk metallicity of [Fe/H] = 0.22 dex. Furthermore, these results are also replicated with the set of $Y^{2}$ parameters. This suggests that the difference in lithium between the two stars cannot be solely explained by differences in parameters.  If this were the case, we would expect Kronos to be defficient in Li compared to Krios, considering Li < 1 dex, following the trend of the metal-rich stars in the sample; however, it has Li=2.78 dex, which is far from this sequence.
 
Due to the considerable depletion of lithium in stars, which can exceed a factor of 100 at the solar age \citep[e.g.][]{asplund09,monroe13}, planet engulfment provides a viable mechanism for significantly increasing the photospheric lithium content in solar-type stars \citep[e.g.][]{ramirez12,melendez17}. \citet{sandquist02}   showed that planet accretion onto the host star could introduce planet material into the stellar convection zone, thereby modifying surface abundances, especially with respect to lithium.

\citet{melendez17} found an increase in Li in HIP 68468 of approximately 0.6 dex, four times more than expected for a star of its age, attributing this phenomenon to a possible planet ingestion. In a similar work, \citet{galarza21} analysed the binary system HIP 71726--HIP 71737. Their analysis revealed a metallicity difference of $\Delta(Fe/H)\sim0.11$ dex and a lithium disparity of $\sim1.03$ dex between the components. The authors concluded that an engulfment event involving $\sim\ 9.8\ M_{\oplus}$ of rocky material could account for these observed differences. \citet{spina21} analysed the chemical composition of 107 binary systems composed of solar-type stars, finding that those stars with higher [Fe/H] than their companions also exhibited an increase in Li abundance, linking both results to planetary ingestion by these enriched objects. They determined that engulfment events occur with a probability of 20-35\%. Nonetheless, \citet{behmard23} claim that the use of an inhomogeneous sample, the omission of an analysis of abundances with  $T_{C}$, and the fact that some binaries in the sample did not qualify as twins could significantly affect the high rates of engulfment found by \citet{spina21}. Instead, they conducted a more detailed analysis of 36 planet-hosting binaries, of which only 11 systems were considered twins, aiming to detect potential engulfment events. This exploration revealed that engulfment events are rare, with a rate of $\sim$2.9\%. Notably, the study emphasizes that only the Krios--Kronos binary could have experienced a genuine engulfment event. 

\subsection{Searching for planets around Kronos \& Krios}
\label{sec:planet}

To date, there have been no planets detected in orbit around Kronos and Krios. Therefore, we conducted a detailed photometric analysis with the aim of revealing potential planetary bodies that could offer valuable insights to the planet formation scenarios expounded in the subsequent sections.

Both stars were observed by the Transiting Exoplanet Survey Satellite mission \citep[TESS;][]{ricker15} in sectors 17, 18, and 24 (from October 8 to November 27, 2019, and from April 16 to May 12, 2020) with a 30-minute cadence and in sectors 57 and 58 (September 30--November 26, 2022) with a cadence of 200 seconds. The analysis of these data products, available in target pixel file (TPF) format, was carried out with the tools provided by the Lightkurve Python package \citep{cardoso18}. Given that both stars are sufficiently separated in the TESS field, we were able to analyse the TPF files of Kronos and Krios independently. We performed single-aperture photometry on the images, choosing as optimal aperture the one centred on the target that allowed   all the possible flux to be collected from the star, but that  minimized the sky contribution. The 30-minute and 200-second cadence light curves were treated separately. For both modes, a median filter was applied to remove the systematics in the resulting light curves. We were not able to eliminate the strong systematics introduced by the changes in the Earth-Moon orientation and distance in sector 24 and, hence, these data were not used in the further analysis.

\begin{figure}[htbp]
    \centering
    \begin{minipage}{0.50\textwidth}
        \centering
        \includegraphics[width=\linewidth]{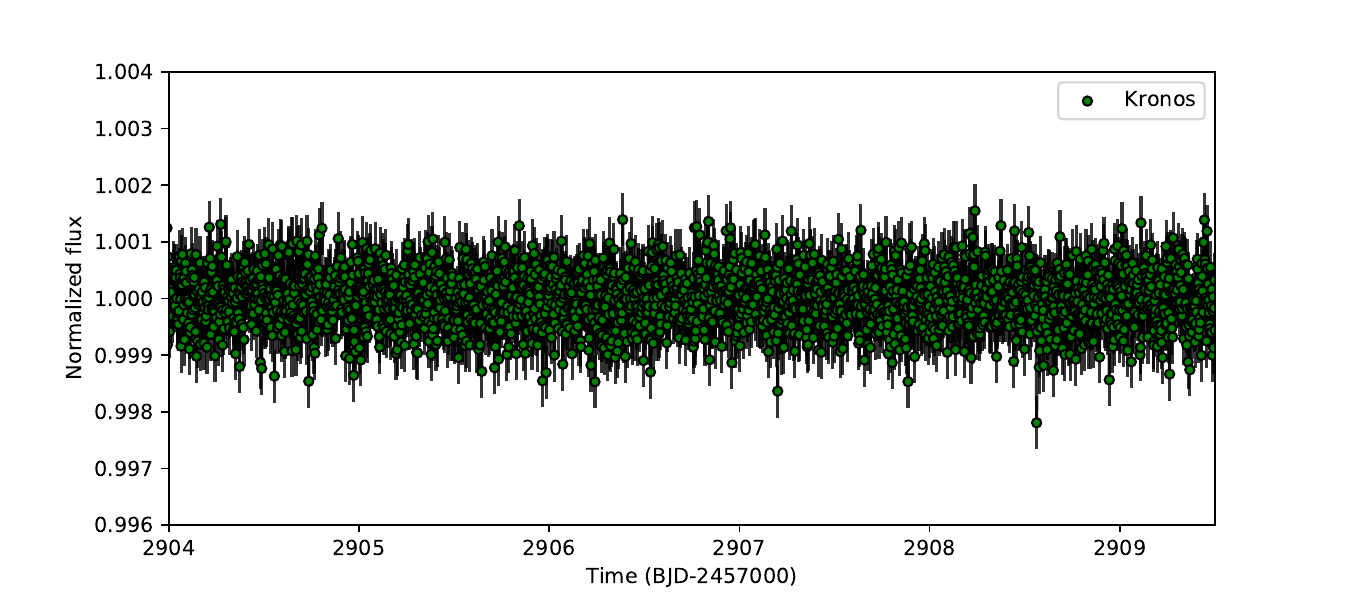}
    \end{minipage}
    \hfill
    \begin{minipage}{0.50\textwidth}
        \centering
        \includegraphics[width=\linewidth]{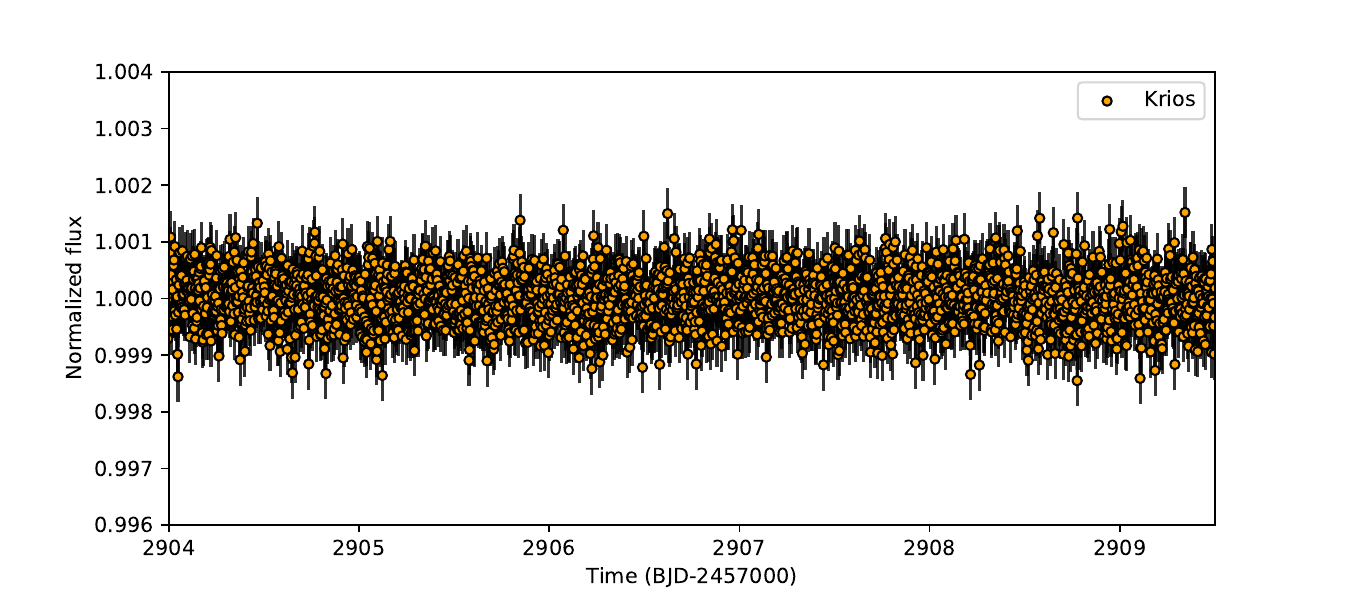}
    \end{minipage}
    \caption{Portion of the detrended TESS light curves of Kronos (top) and Krios (bottom) considering the 200-second cadence data of sector 58.}
    \label{curvas}
\end{figure}

To look for signs of additional stellar and/or planetary companions around Kronos and Krios, we ran the Transit Least Squares code \citep[TLS;][]{hippke19} on the detrended light curves of each component separately (Figure \ref{curvas}). No transit or eclipse-like signal that could suggest the presence of a transiting
planet or an eclipsing stellar companion was detected in the 30-minute or in the 200-second cadence of the two stars. Additionally, a detailed by-eye inspection of the TESS photometry revealed that the stars show no signs of periodic modulation or sporadic events, such as flares, which indicates that they are not photometrically active objects. Here, it is important to caution that the present conclusion about the periodic photometric variability is based only on visual scrutiny of the data. In order to obtain a more reliable and confident result,  we should run on the TESS light curves of Kronos and Krios a tool specifically designed to detect periodic modulations in time series, such as the Lomb–Scargle periodogram \citep{lomb76,scargle82} or the auto-correlation function \citep{mcquillan13}. However, conducting such an analysis is beyond the scope of this paper.

\subsection{Atomic diffusion}

The atomic diffusion process includes effects such as gravitational settling, thermal and chemical diffusion, and radiative acceleration \citep[e.g.][]{dotter17, liu21}. 
It primarily operates in the radiative zones of the stars, pushing certain elements and altering its surface abundances, depending on the particular species and the evolutionary state of the star. In the case of substantial differences in the spectroscopic parameters of stars ($T_{eff}$ or $log\ \textit{g}$) forming a binary system, this process could potentially explain a disparity in metallicity between the components since their abundances may have been affected differently as they evolve. \citet{liu21} found that, for four of the seven studied pairs with differences in log\ \textit{g} $>0.05$ dex, their discrepancies in [Fe/H] could be attributed to atomic diffusion rather than planetary formation.

\begin{figure}
\centering
\includegraphics[width=9cm]{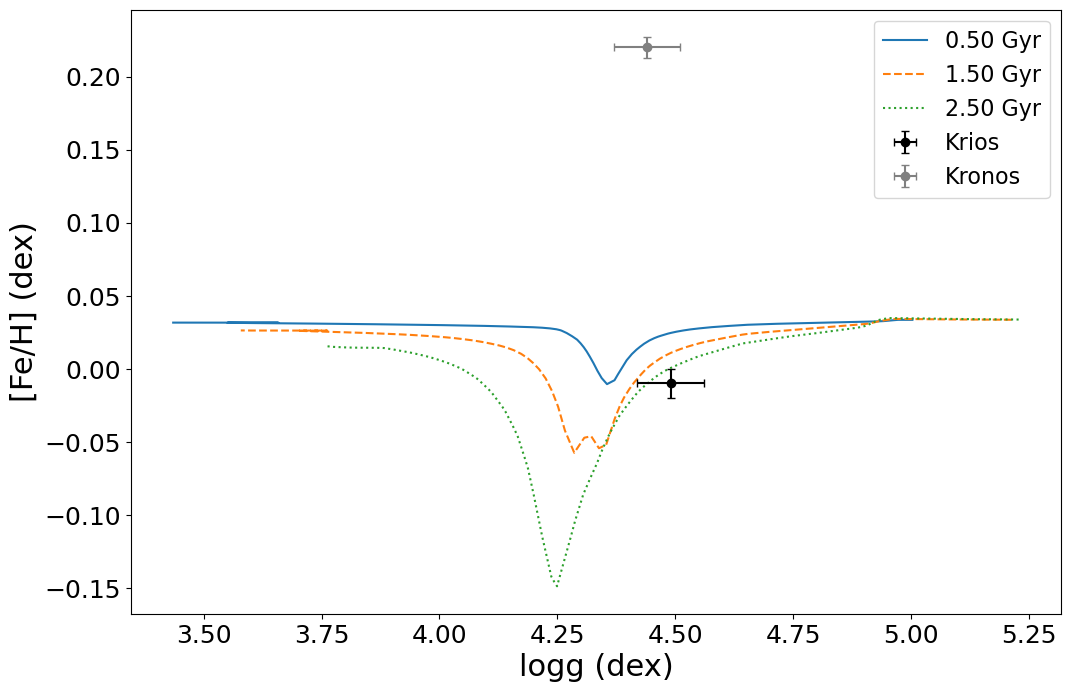}
\captionsetup{justification=justified,singlelinecheck=false} 
\caption{Set of isochrones for an age range of 0.5--2.5 Gyr. Krios is plotted in black and Kronos in grey. The vertical and horizontal bars correspond to $\sigma[Fe/H]$ and $\sigma\ log\ \textit{g}$, respectively.}
\label{difusion}%
\end{figure}

In the present work, Kronos and Krios exhibit a log\ \textit{g} difference of approximately 0.05 dex. In consequence, we investigate whether the observed difference in metallicity between the components could be attributed to a diffusion process. To address this, we used the MESA Isochrones and Stellar Tracks \citep[MIST;\footnote{\tiny \url{https://waps.cfa.harvard.edu/MIST/index.html}}][]{choi16}, which facilitates the derivation of stellar evolutionary models that integrate the influences of atomic diffusion and overshoot mixing, and also employed solar abundances from \citet{asplund09}. We generated a set of isochrones covering the age range of both stars; the results are depicted in Figure \ref{difusion}. From the figure, it is evident that Krios follows an evolutionary model consistent, within the errors, with the age calculated from PARAM ($\tau_{\star}\approx1.57 \ Gyr$). With Kronos exhibiting a lower log\ \textit{g} than Krios, the metallicity of Kronos would be expected to be lower if diffusion was predominant in explaining the anomalies found. However, as depicted in Figure \ref{difusion}, Kronos exhibits a significantly higher metallicity than Krios. Based on this result, while this effect cannot be completely ruled out, we can consider that it is not the primary factor responsible for the pronounced difference in metallicity found in the binary system. There must be an additional mechanism to account for these discrepancies.

\subsection{Primordial chemical differences between components}

Binary systems are ideal laboratories for testing a number of scenarios that have been proposed to explain the origin of chemical signatures. This is attributed to the shared origin of the two stars within the same molecular cloud, assuming that their primordial chemical composition should be similar and diminishing the factor of GCE.

Taking into account the substantial difference in metallicity between Kronos and Krios, having a projected separation of approximately 11277 au, \citet{oh18} explored the probability of their coincidental pairing. Using the \textit{Gaia} Universe Mock Simulation \citep{robin12} and the Besançon Galaxy model \citep{robin03}, they looked for chance pairs within 200 parsec of the Sun. From a sample of 119259 solar-mass primary stars, they  found only one pair with $\Delta v_{r}<2\  km\ s^{-1}$, which naturally suggests a physical association between the Kronos \& Krios system rather than a chance pairing. We further calculated the $\Delta v_{\rm 3D}$ of the binary system. For this purpose, we utilized the space velocities of each component from the \textit{Gaia} DR3 dataset  \citep{gaia21} using the \textsc{Gala} code \citep{price17}. The $\Delta v_{\rm 3D}$ is estimated at $0.55\ km\ s^{-1}$, which is below the $2\ km\ s^{-1}$ limit required to ensure the continuity of a binary system \citep{kamdar192}.

        \citet{ramirez19} examined a sample of 12 binary systems with twin stars, and found a modest correlation between the absolute difference in metallicity among the components and their separation. They found increased metallicity discrepancies with expanding separations between the binary system components. Additionally, \citet{andrews19} investigated chemical homogeneity in 24 binary systems with similar components, finding consistency in their abundances at a level of 0.1 dex. Furthermore, they generated a set of random pairs from these systems, and in this case, consistency was observed at a level of 0.3--0.4 dex. Following this line, \citet{nelson21} analysed 33 comoving pairs of F and G dwarfs and found that those comoving systems spanning separations from $\sim 2 \times 10^{5}$ au to $2 \times 10^{7}$ au exhibit greater homogeneity ($\Delta[Fe/H]=0.09$ dex) than those that are randomly paired ($\Delta[Fe/H]=0.23$ dex). 
        
        Assuming that the two stars indeed constitute a coeval and conatal system, and considering the results presented by \citet{nelson21}, it suggests that the difference in metallicity found in the binary system cannot be solely explained by their separation distance, suggesting the existence of some other factor to account for this significant discrepancy. 
        
\subsection{Planet formation scenarios}
\label{sec:planet_formation}

\subsubsection{Rocky planet formation}

The average abundance calculated for refractory and volatile elements for Krios-Kronos are $-0.24 \pm 0.01$ dex and $-0.03 \pm 0.03$ dex, respectively. These results,   along with the trend observed in Figure \ref{tcond}, indicate an overabundance of refractories in Kronos compared to Krios. Moreover, as seen in Section \ref{sec:Li}, there is an excess of Li in Kronos of $\Delta (Li) = 0.56\ $dex, which cannot be solely explained by differences in the parameters of the components.

Among the possible scenarios that could explain this result, we firstly consider the hypothesis presented by \citet{melendez09}. They suggested that the lack of refractories in the Sun may be attributed to the formation of terrestrial planets and planetesimals around it, which primarily accreted refractory material for this purpose \citep[e.g.][]{saffe16,yana16-2,liu20}. The fact that Krios exhibits a deficiency in refractories could potentially result from a protoplanetary disc sequestering refractory material, possibly for the subsequent formation of rocky planets around the star. To date, as we present  in Section \ref{sec:planet}, no planets have been detected transiting either component of the binary system. While current evidence does not provide strong support for this model, it would be intriguing to conduct a radial velocity study to search for possible anomalies that could indicate the presence of a planet.

Another plausible factor that may account for the disparity in metallicity and the trend with $T_{C}$ among the binary system components is the potential presence of a debris disc encircling Krios, similar to what was observed in the $\zeta^{1}-\zeta^{2}$ Ret system \citep{saffe16}. Debris disc detection primarily relies on identifying infrared (IR) excess emissions originating from circumstellar dust. These dust particles have lifespans shorter than those of stellar systems, reinforcing the hypothesis that these discs experience continuous replenishment through ongoing collisions with substantial celestial bodies \citep[e.g.][]{wyatt08}.

To investigate the potential presence of an IR excess in this binary system, we employed the VOSA\footnote{\tiny \url{http://svo2.cab.inta-csic.es/theory/vosa/index.php}} platform, obtaining the energy distribution for both system components using photometric observations from the SDSS, JPAS, TYCHO, JPLUS, Johnson, WISE, 2MASS, and GAIA3 filters. The analysis did not reveal any IR excess in either component, discouraging the possibility that the observed differences in metallicity and the $T_{C}$ trend in this system are linked to the presence of a debris disc around Krios.

\subsubsection{Dust trapping}

The model proposed by \citet{booth-owen20} suggests that the lack of refractories in one of the stars could be due to the formation of a massive gas giant planet that created a gas gap in the protoplanetary disc. This gap transforms into an outer pressure trap beyond the orbit of the planet, mainly sequestering dust from the disc. This mechanism could reveal a disparity between refractory and volatile elements in the hosting star. The recent study by \citet{huhn23} refines the understanding of this planetary formation scenario, exploring how the origin of a planet influences the material accreted onto the convective envelope. 
 
 If we consider this scenario as plausible, Krios should have a Jupiter-sized planet in orbit, which, according to this model, would create traps allowing the accretion of volatiles while inhibiting the accretion of refractories. If this were the case, it could potentially account for the abundance pattern observed in Figure \ref{tcond}. The absence of detected planets to date does not provide conclusive evidence to entirely rule out this hypothesis.

\subsubsection{Planetary ingestion}

Another important scenario to consider is the engulfment hypothesis \citep[e.g.][]{saffe17,galarza21,jofre21, flores23}. \citet{spina21} suggested that two conditions must be met for the observed anomalies to be attributed to the engulfment of a planet. Firstly, there should be an excess of refractory elements compared to volatiles in one of the stars in the pair, indicating the accretion of rocky material by that object. Secondly, it should also exhibit an excess of Li compared to its companion. This latter characteristic becomes particularly significant when engulfment occurs at an advanced age of the star because, by that time, it would have already burned most of the Li in its atmosphere, and therefore the accretion of new refractory material would leave a substantial and detectable imprint when comparing the A(Li) of the two components.

 When comparing our findings with the hypotheses presented in the work of \citet{spina21}, the scenario of engulfment becomes a plausible consideration. This implies that Kronos might have undergone the accretion of one or more planets at an advanced age, leaving distinctive lithium content marks and introducing refractory elements into the atmosphere of the star.  To determine how much terrestrial mass Kronos would need to have accreted to achieve these values, we employed the \textsc{terra} code\footnote{\url{https://github.com/ramstojh/terra}} \citep{yana16}. Our estimations reveal a convection envelope mass of $M_{cz}= 0.017M_{\odot}$ for Kronos, with approximately $\sim27.8M_{\oplus}$ of rocky material considered necessary to replicate the observed trend illustrated in Figure \ref{tcond}. This includes a combination of $19.9M_{\oplus}$ of terrestrial material and $7.9M_{\oplus}$ of meteoritic material. This remarkable magnitude of ingested material represents one of the largest estimated to date in twin components, underscoring the significance of the results. Furthermore, recent research conducted by \citet{armstrong20} provides compelling evidence of the existence of TOI849-b, a planet with a core mass of $39.1M_{\oplus}$. This discovery not only reinforces the plausibility of our findings, but also highlights the prevalence of planetary bodies with masses comparable to or even greater than the values found in our study within exoplanetary systems.
 
\begin{figure}
\centering
\includegraphics[width=9.5cm]{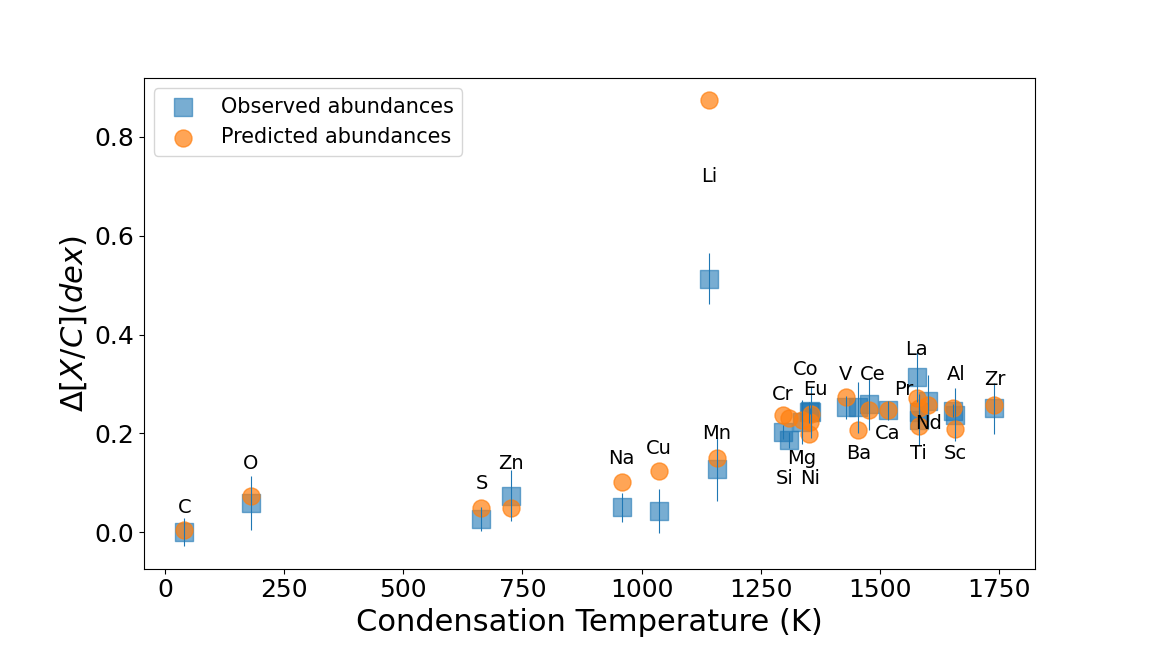}
\captionsetup{justification=justified,singlelinecheck=false} 
\caption{Observed and predicted abundances of Kronos--Krios vs $T_{C}$ (blue squares and orange dots, respectively), considering that Kronos ingested $\sim27.8M_{\oplus}$.}
\label{modelo}%
\end{figure}

Figure \ref{modelo} depicts the observed model and the model adjusted by \textsc{terra}, taking into account the ingestion of $\sim27.8M_{\oplus}$ . A good agreement between the two models is evident, except for Li, which is overestimated by $\sim0.36$ dex. We caution that \textsc{terra} models engulfment as if it had occurred at the present time. Hence, the excess of Li found in the predicted model suggests that the engulfment likely occurred in the past. The mass of the convective envelope plays a significant role; therefore, knowing this value at the time of accretion would contribute to improving the model. Nonetheless, the simulation predicted by \textsc{terra} can be considered   a solid first approximation.

\citet{behmard23} employed a sample of 36 stellar systems to quantify the duration a chemical signature could remain observable in the stellar photosphere due to the ingestion of a planet and its associated strength. They simulated pollution resulting from the engulfment of a 10 $M_{\oplus}$ planet and found that stars with masses ranging between 1.1-1.2 $M_{\odot}$ exhibited the highest and most enduring chemical signature, maintaining values greater than 0.05 dex for approximately 2 Gyr. They also conducted an analysis considering the ingestion of a 50 $M_{\oplus}$ planet, where stars with masses between 0.7-1.2 $M_{\odot}$ displayed signatures exceeding 0.05 dex for a duration spanning 3-8 Gyr. Taking into account the age and mass of Kronos, these findings lead us to consider that the chemical differences observed in this star when compared to Krios, may have originated from a potential engulfment of rocky material.

The Kozai migration \citep{kozai62} has been proposed to explain this phenomenon in other binary systems, wherein a giant planet orbiting one of the stars may experience orbital decay due to a combination of perturbations caused by the other star in the system that leads to an increase in the eccentricity of the planetary orbit, accompanied by tidal friction that brings the planet closer to the host star, ultimately resulting in the ingestion of the surrounding rocky material and potentially the planet itself \citep{wu03,takeda08,borkovits11,mustill15,petrovich15,church20}. In this context, we should assume that Kronos initialy formed a giant gas planet as well and possibly rocky material. Then the migration of this hypothetical planet triggered the accretion of refractory material, either from the inner regions of the planetary system or from the giant planet’s core itself, resulting in the observed refractory excess (as shown in Fig. \ref{tcond}). Some similar migration scenarios have been invoked in the literature for other binary systems \citep[e.g.][] {neveu14, teske15, saffe17, jofre21, flores23}.

\section{Conclusions}

We performed a high-precision differential abundance analysis of the binary system Krios \& Kronos with the aim of exploring different scenarios that could explain the particularly high [Fe/H] disparity found in \citet{oh18}. To achieve this, we took advantage of high-resolution spectra (S/N $\sim 420$) obtained from MAROON-X. We calculated the fundamental atmospheric parameters ($T_{eff}$, log\ \textit{g}, [Fe/H], $v_{turb}$) for the two stars,  for the first time making
use of the non-solar-scaled method and  using the Sun as reference, and recalculated parameters of Krios using Kronos as reference. We also measured chemical abundances for 27 elements through equivalent widths and spectral synthesis, subsequently analysing their relation with the $T_{C}$. We found high similarity in the fundamental parameters of the two components and confirmed the existing difference in metallicity between them, with Kronos having a metallicity $\sim 0.230$ dex higher than Krios. This substantial disparity suggests that previous chemical tagging works may not have successfully recovered their shared origin  \citep[e.g.][]{desilva06,bovy16,liu16,casamiquela20,casamiquela21}. 

In addition to these results, a significant difference in Li abundance between the components was also found, with Kronos being 0.56 dex more abundant in Li than Krios. When comparing the abundances of (Krios-Kronos) versus $T_{C}$, we observed a pronounced trend relative to $T_{C}$, a behaviour that is repeated when considering only refractory elements. From these results, we primarily deduce an excess of refractories in Kronos compared to Krios. 

We conducted a comprehensive single-aperture photometry analysis using TESS data and the TLS code to investigate potential planets orbiting either of the stars. No transits or eclipses of potential planets orbiting any of the components were detected, and there were no indications of stellar activity. While no evidence of transiting planets around Kronos and Krios was found, it should be noted that planetary-mass bodies that do not transit may still exist in the system. Additionally, it would be compelling to perform an analysis of the radial velocity variations of both components to shed light to this hypothesis.

Different scenarios were considered to explain the results obtained. We introduced, for the first time, an atomic diffusion analysis in this system, given the 0.05 dex difference in log\ \textit{g} found between components, the limit considered by \citet{liu21} beyond which this phenomenon could affect the metallicity of the components. However, the characteristics of Kronos differ significantly from what was anticipated by its evolutionary model, suggesting that this scenario may not entirely account for the wide difference in metallicity.

We also examined the potential that this difference had a primordial origin, considering the projected separation existing between the stars ($\sim11277$ au). Following the approach of \citet{nelson21}, given that comoving pairs exhibit differences in metallicity of $\Delta[Fe/H]\sim0.09\ $dex, and taking into account that both stars probably formed from the same gas and dust cloud \citep{oh18}, we suggest that the difference in metallicity between the components cannot be solely associated with primordial differences; this implies the presence of an additional factor influencing this substantial disparity.

Planet formation scenarios were also investigated. The $ T_{C}$ trend found in Figure \ref{tcond}, if interpreted as a deficiency of refractories in Krios, could have its origin in the formation of rocky planets (not yet detected), as proposed by \citet{melendez09}. We also analysed the IR excess, searching for a possible dust disc in Krios that could be generating the observed effect, with no positive results. Additionally, the scenario proposed by \citet{booth-owen20} was analysed in this binary system for the first time, assuming the presence of a hypothetical Jupiter-sized planet orbiting Krios, in which case pressure traps that sequester refractory elements could generate the observed pattern. However, additional photometric and spectroscopic data are necessary to conduct a more detailed search for planets around Krios, and to shed light on this hypotheses.

The last scenario analysed was planetary engulfment, a phenomenon whose characteristics closely match the results found in Kronos, both in the excess of [Fe/H] and the excess of Li compared to Krios \citep{spina21}. Regarding this hypothesis, we calculated the amount of rocky material Kronos would have ingested to achieve this difference through the \textsc{terra} code, resulting in $\sim27.8M_{\oplus}$. 

In conclusion, while the evidence may seem to favour the engulfment hypothesis, it is crucial to acknowledge the inherent complexities and uncertainties associated with each scenario. Therefore, further investigation and exploration are imperative in order to  achieve a more comprehensive understanding of the chemical anomalies and dynamics within this binary system.

\section*{Acknowledgments}

PM and JA acknowledge Consejo Nacional de Investigaciones Científicas y Técnicas (CONICET) for their financial support in the form of doctoral fellowships. R.P. acknowledges funding from CONICET, under project PIBAA-CONICET ID-73811. We acknowledge the use of public TESS data from pipelines at the TESS Science Office and at the TESS Science Processing Operations Center. E.J. acknowledges funding from CONICET, under project number PIBAA-CONICET ID-73669. JYG acknowledges the generous support of a Carnegie Fellowship. 
This work was enabled by observations made from the Gemini North telescope, located within the Maunakea Science Reserve and adjacent to the summit of Maunakea. We are grateful for the privilege of observing the Universe from a place that is unique in both its astronomical quality and its cultural significance.
The international Gemini Observatory, a programme of NSF’s NOIRLab, is managed by the Association of Universities for Research in Astronomy (AURA) under a cooperative agreement with the National Science Foundation on behalf of the Gemini partnership: the National Science Foundation (United States), the National Research Council (Canada), Agencia Nacional de Investigación y Desarrollo (Chile), Ministerio de Ciencia, Tecnología e Innovación (Argentina), Ministério da Ciência, Tecnologia, Inovações e Comunicações (Brazil), and Korea Astronomy and Space Science Institute (Republic of Korea). The MAROON-X team acknowledges funding from the David and Lucile Packard Foundation, the Heising-Simons Foundation, the Gordon and Betty Moore Foundation, the Gemini Observatory, the NSF (award number 2108465), and NASA (grant number 80NSSC22K0117). 

%
%

\end{document}